# Imaginary Gauge Fields for One-Way Transparency and Absorption in a Passive Metasurface

Qingdong Yang [1], Zhongfu Li[1], Xinhua Wen [1,2], Oubo You[1], Yi Wang[3&], Shuang Zhang[1,4,5,6,7]*

**Affiliations:**

[1] New Cornerstone Science Laboratory, Department of Physics, University of Hong Kong; 999077, Hong Kong, China

[2] School of Physics and Optoelectronics, South China University of Technology, Guangzhou, Guangdong 510640, China

[3] School of Engineering, University of Birmingham, Birmingham B15 2TT, U.K

[4] State Key Laboratory of Optical Quantum Materials, University of Hong Kong, Hong Kong 999077, China

[5] Quantum Science Center of Guangdong-Hong Kong-Macao Greater Bay Area, 3 Binlang Road, Shenzhen 518000, China

[6] Materials Innovation Institute for Life Sciences and Energy (MILES), HKU-SIRI, Shenzhen, China

[7] Department of Electrical & Electronic Engineering, University of Hong Kong; 999077, Hong Kong, China.

[&]Corresponding author. Email: y.wang.1@bham.ac.uk

*Corresponding author. Email: shuzhang@hku.hk

## Abstract

Electromagnetic nonreciprocity enables waves to respond differently when their propagation direction is reversed, forming the basis of isolation, directional routing, and asymmetric energy control. A central challenge is to achieve high transmission in one direction while inducing strong absorption in the opposite direction within a single passive element, as passive material dissipation typically attenuates both propagation channels equally. Here we demonstrate that an imaginary artificial gauge field can redistribute net dissipation between opposite directions in a passive structure. By synthesizing a moving-type magnetoelectric response from gyromagnetic elements and subwavelength metallic resonators, we realize a polarization-independent metasurface in which the forward wave weakly excites the dissipative resonance through destructive current interference, whereas the backward wave strongly activates the same lossy mode through constructive interference. The fabricated metasurface transmits more than 80% of the incident power from one side while absorbing more than 80% from the opposite side, with low reflection from both directions. Near-field mapping of the surface electric field provides direct real-space evidence of this gauge-controlled, direction-dependent charge accumulation and dissipation. This work establishes imaginary gauge fields as a powerful route for engineering dissipative landscapes in open wave systems and opens a pathway toward compact, passive, reflectionless isolators and nonreciprocal absorbers.

# Main

Optical nonreciprocity breaks the fundamental time-reversal symmetry of wave propagation, allowing electromagnetic waves to respond differently when their propagation direction is reversed[1]. This unidirectional control is essential for vital photonic applications, including optical isolation, routing, phase control, directional emission, and topological transport[2-8]. Traditional nonreciprocal devices, such as conventional isolators, typically achieve macroscopic one-way transport by combining magneto-optical effects[9-11], nonlinear optical responses[12-15], and spatiotemporal modulation[16,17]. However, these systems usually operate at the system level: the unwanted wave is separated from the desired channel through polarization filtering, port redirection, mode conversion, diffraction into auxiliary channels, or multi-component integration before being dissipated[18,19]. As a result, they often introduce additional constraints, such as long interaction lengths, polarization sensitivity, power-dependent operation, sideband generation, or increased system complexity[20].

A long-standing frontier in photonics is to realize simultaneous high transmission in one direction and strong absorption in the opposite direction within a single, passive, and compact element. The central difficulty lies in the nature of material dissipation: in conventional media, loss attenuates waves equally from both directions, making it fundamentally incompatible with one-way transparency. Reciprocal metasurfaces can achieve asymmetric absorption through structural asymmetry, but this typically comes at the cost of high reflection from one side[21,22]. True reflectionless nonreciprocal transmission combined with directional absorption in a passive structure has remained elusive.

Artificial gauge fields provide a powerful route for controlling wave propagation beyond conventional refractive-index and impedance engineering.[23-26] In photonics, gauge potentials have been used to engineer synthetic magnetic fields, nonreciprocal phase accumulation, and robust wave routing[27-32]. However, their potential for manipulating dissipation has been largely unexplored. Here we show that an imaginary artificial gauge field can transform intrinsic material loss into a directional design degree of freedom. By introducing a complex gauge potential, the net attenuation accumulated along opposite propagation paths becomes unequal. The imaginary component of the gauge field redistributes dissipation: it suppresses loss for forward propagation while enhancing it for backward propagation, all within a passive system.

Here we experimentally demonstrate this concept through a passive, polarization-independent metasurface operating at microwave frequencies. The structure synthesizes a moving-type magnetoelectric (bianisotropic) response[33-35] by integrating gyromagnetic yttrium iron garnet (YIG) rods with subwavelength metallic resonators. This response implements the required imaginary gauge potential, leading to direction-dependent excitation of the same dissipative resonant mode—destructive current interference for forward incidence and constructive interference for backward incidence. The fabricated metasurface transmits more than 80% of the incident power from the forward direction while absorbing more than 80% from the reverse direction, with reflections from both sides strongly suppressed. Near-field mapping directly visualizes the gauge-controlled charge accumulation that underlies this behavior. Our work establishes imaginary gauge fields as a versatile platform for engineering direction-dependent dissipation in open wave systems. Beyond providing a route toward compact, passive, reflectionless isolators and asymmetric absorbers, it opens new opportunities in non-Hermitian wave physics and gauge-regulated loss landscapes.

## Artificial-gauge design for direction-dependent loss

We begin by formulating the required response using a simple propagation picture for an $E_x$-polarized wave traveling along the $z$ direction. In an ordinary medium, the Maxwell equation takes the form $\partial_z \psi = i k_0 \begin{pmatrix} 0 & \mu \\ \varepsilon & 0 \end{pmatrix} \psi$, where $\psi = \left(E_x, \eta_0 H_y\right)^T$, where $k_0, \eta_0$ are vacuum wavenumber and impedance, respectively. The propagation constant $n_0 = \sqrt{\varepsilon\mu} = n' + in''$. Here, $k_0 n'$ determines the propagation phase, whereas $k_0 n''$ gives the passive attenuation. Because this attenuation depends only on the propagation length, it is identical for waves traveling in the forward and backward directions. Conventional material loss therefore cannot produce direction-dependent behavior. To achieve direction-selective attenuation, we introduce a uniform artificial gauge potential $A_z$ through the covariant derivative as shown in Fig. 1. The propagation equation becomes

$$(\partial_z + iA_z)\psi = i k_0 \begin{pmatrix} 0 & \mu \\ \varepsilon & 0 \end{pmatrix} \psi \tag{1}$$

The effect of the gauge potential is most clearly seen as a Wilson-line-like propagation factor accumulated across the metasurface[36] (Supplementary Materials Sec. 1). It therefore contributes through a path-dependent phase factor to original wave propagation. For forward propagating wave, it contributes an extra gauge factor $\exp\left(-i \int_0^d A_z dz\right) = \exp(-iA_z d)$, where $d$ is the propagation distance. Decomposing the complex gauge potential as $A_z = A_z' + iA_z''$, the imaginary component $A_z''$ modifies the effective attenuation. The forward attenuation becomes

$$\alpha_+ = k_0 n'' - A_z''. \tag{2}$$

While for the backward-propagating wave, the sign of the gauge accumulation reverses, yielding

$$\alpha_- = k_0 n'' + A_z''. \tag{3}$$

Thus, the ordinary material loss $k_0 n''$ affects both directions equally, whereas the imaginary gauge field $A_z''$ redistributes the attenuation between the two directions: increasing $A_z''$ suppresses loss in the forward direction while enhancing it in the backward direction. When $A_z'' \approx k_0 n''$, the forward channel approaches transparency, while the backward channel is driven toward strong absorption.

**Physical implementation of the dissipative artificial gauge field**

Here we show that, in the effective-medium picture, the moving-type magnetoelectric coupling enters the wave equation in the same form as a gauge potential. Its constitutive equation is given by

$$\begin{pmatrix} D_x \\ D_y \end{pmatrix} = \varepsilon_0 \varepsilon \begin{pmatrix} E_x \\ E_y \end{pmatrix} + \frac{1}{c_0} \begin{pmatrix} 0 & -v \\ v & 0 \end{pmatrix} \begin{pmatrix} H_x \\ H_y \end{pmatrix}$$

$$\begin{pmatrix} B_x \\ B_y \end{pmatrix} = \mu_0 \mu \begin{pmatrix} H_x \\ H_y \end{pmatrix} + \frac{1}{c_0} \begin{pmatrix} 0 & v \\ -v & 0 \end{pmatrix} \begin{pmatrix} E_x \\ E_y \end{pmatrix} \tag{4}$$

where $v$ is the effective velocity of the moving medium[37,38]. It can be shown that Eqn. 4 can be transformed into Eq. 1 with $v = A_z / k_0$ (see details in Supplementary Materials Sec. 2). This establishes the link between the effective gauge description and the physical metasurface response. Its real part modifies the propagation phase, whereas its imaginary part acts as the dissipative gauge component that redistributes attenuation between forward and backward waves.

To realize this response, we design a polarization-independent metasurface whose unit cell is shown in Fig. 2a. Each unit cell consists of four metamolecules, each containing an YIG rod centered within a metallic resonator and biased by permanent magnets. The metallic resonator is formed by two planar helical patterns with opposite handedness. By carefully arranging the local magnetic bias and the resonator geometry, the moving-type magnetoelectric responses from different metamolecules add constructively, while unwanted gyromagnetic and chiral responses are suppressed (see details in Supplementary Materials Sec. 3). The resulting checkerboard arrangement preserves a fourfold rotational symmetry and in-plane mirror symmetry, ensuring polarization-independent scattering and absorption.[39]

Having established $A_z''$ as the effective parameter controlling direction-dependent loss, we now examine the microscopic origin in terms of dissipative current interference. Because the operating frequency lies near the structural resonance of the metallic pattern, the electromagnetic response is dominated by strong circulating and displacement currents. These currents strongly confine the local fields, generate the effective electric and magnetic dipoles, and provide the primary channel for dissipative excitation. Taking one metamolecule in Fig. 2b as an example, for forward incidence, the incident electric field $E_x$ directly drives charge accumulation at the two ends of the metallic arms, producing a resonant current $J_E$ (pink arrows)[40]. Simultaneously, the magnetic field $H_y$ excites magnetic dipoles in YIG rods along x-axis due to the gyromagnetic effects, which in turn induces an electromotive force in the nearby metallic resonator via Faraday induction. This generates a secondary charge accumulation and current $J_M$ (green arrows). The phase lags from the gyromagnetic response of YIG ($\pi/2$) and the Faraday-induction process ($\pi/2$) cause $J_M$ to oppose $J_E$ in the critical resonant arms. The two currents interfere destructively, reducing the total current amplitude, $J_{\text{tot}}^{(+)} \approx J_E - J_M$, weakening the field hot spots, and suppressing the resonant dissipation. Consequently, the metasurface remains nearly transparent from the forward side.

For backward incidence (Fig. 2c), the directly driven current $J_E$ retains the same reference phase. However, reversal of the propagation direction flips the sign of $H_y$, which reverses the induced current $J_M$. The two contributions now add constructively ($J_{\mathrm{tot}}^{(-)} \approx J_E + J_M$), producing stronger hot spots and enhanced dissipation. This direction-dependent activation of the same lossy resonant current mode is the microscopic manifestation of the imaginary gauge field: the forward gauge path suppresses the dissipative excitation, whereas the reversed path enhances it. Although the current interference picture mostly reveals the mechanism, the total absorption also includes contributions from dielectric and gyromagnetic losses. A full dissipated-power analysis is provided in Supplementary Materials Sec. 4.

We next verify that the gauge-controlled redistribution of loss produces the desired far-field response. The co-polarized transmission and reflection spectra are calculated for plane waves incident from both sides of the metasurface as shown in Fig. 3. The results show a strong direction-dependent transmission near 9.3 GHz. For forward incidence, the metasurface exhibits high transmission, with $|t_+|^2$ reaching approximately 85%. For backward incidence, the transmission is strongly suppressed, with $|t_-|^2$ decreasing to 5%. At the same frequency, the reflection remains small for both incident directions. This results in backward absorbance exceeding 90%, as shown in Fig. 3b.

To connect these scattering properties to the artificial gauge picture, we retrieve the effective parameters by solving Fresnel's equation from the simulated complex transmission and reflection coefficients (refer to Supplementary Materials Sec. 5). Figure 3c presents the imaginary part of refractive index $n''$, representing the common passive loss background, and the imaginary part of the moving coupling parameter $v''$ (the dissipative gauge component). The retrieved parameters satisfy the passivity constraints and the physical constraints on magnetoelectric coupling.[41,42] At the operating frequency, the retrieved $v''$ approaches the ordinary loss background $n''$ ($v'' \approx n''$), nearly canceling attenuation for the forward wave while doubling it for the backward wave. Away from the operating frequency, for example near 9.5 GHz, $v''$ changes sign. This sign reversal does not indicate the emergence of gain; rather, it corresponds to a reversal of the effective imaginary gauge field. Consequently, the direction of loss redistribution is reversed: the forward attenuation becomes enhanced, whereas the backward attenuation is reduced. Since this reversed compensation is incomplete, the transmission contrast is reduced. These retrieved parameters thus confirm that the imaginary gauge field is responsible for the observed nonreciprocal transmission and absorption.

**Experimental demonstration of simultaneous nonreciprocal transmission and absorption**

We fabricate the metasurface using standard printed circuit board technology, with details provided in the Supplementary Materials Sec. 6. Far field transmission and reflection spectra are measured in the microwave regime using a pair of linearly polarized wideband horn antennas connected to a vector network analyzer. The experimental results (Fig. 3d) show excellent agreement with full-wave simulations. At the operating frequency of 9.3 GHz, the forward transmittance reaches 83%, while the backward transmittance drops to 5%. Reflection remains low from both sides, yielding a backward absorbance of 85% (Fig. 3e). The small discrepancy between simulation and experiment can be attributed to fabrication tolerances, nonuniform magnetic bias inside the YIG rods, and finite-size effects of the sample. Effective parameter retrieval from the measured scattering coefficients (Fig. 3f) confirms that the imaginary parts of the refractive index and the moving-type magnetoelectric coupling parameter closely match the simulated values. This verifies that the imaginary gauge field successfully redistributes dissipation in the fabricated structure, enabling one-way transparency combined with strong absorption.

**Near-field evidence of gauge-controlled dissipative excitation**

To directly visualize the action of imaginary gauge field inside the metasurface, we map the out-of-plane electric field $E_z$ on the surface, which serves as a near-field proxy for resonant charge accumulation. Since $E_z$ is directly linked to the surface charge density of the dominant resonant current mode, its spatial distribution reveals how strongly the lossy resonance is excited from each side.

Full-wave simulations show clear direction-dependent behavior at the operating frequency (Fig. 4a, b). For forward incidence, the directly driven current $J_E$ and the gauge-induced $J_M$ interfere destructively, producing only weak $E_z$ hot spots and suppressed dissipation. In contrast, for backward incidence, $J_E$ and $J_M$ add constructively, resulting in strongly enhanced Ez localization near the resonator gaps and metallic arms, indicating intensified resonant currents and higher dissipated power.

Experimentally, we measure the near-field distribution above the metasurface using a scanning dipole probe (See Methods for details). The measured maps (Fig. 4c, d) closely reproduce the simulated contrast: forward incidence only yields weak $E_z$ hot spots around the metallic resonators, consistent with weak excitation of the lossy resonant mode, whereas backward incidence produces pronounced field enhancement. Quantitative amplitude of $E_z$ field is provided in Supplementary Materials Fig. 3. Thus, these real-space measurements provide direct experimental confirmation that the designed passive structure supports a gauge-controlled dissipation.

In summary, we have realized a passive metasurface that achieves simultaneous nonreciprocal transmission and nonreciprocal absorption through gauge-controlled dissipation. The imaginary component of the artificial gauge field, implemented via moving-type magnetoelectric coupling, creates a direction-dependent loss landscape: the forward wave experiences suppressed excitation of the dominant lossy resonance, while the backward wave encounters enhanced dissipation of the same mode. This yields high forward transmittance with low reflection alongside strong backward absorption, all within a compact, polarization-independent structure.

The microscopic origin—direction-dependent interference of resonant currents in the metallic resonators—is directly confirmed by near-field mapping of surface charge accumulation. Effective parameter retrieval further links the observed scattering asymmetry to the imaginary gauge potential that redistributes the common material loss between propagation directions. Unlike reciprocal asymmetric absorbers, which rely on impedance mismatch and typically incur high reflection, our approach maintains low reflection from both sides while breaking reciprocity passively.

This demonstration establishes artificial gauge fields as an effective tool for engineering dissipative environments in open wave systems. It provides a practical route toward compact reflectionless isolators, nonreciprocal thermal emitters, and integrated nonreciprocal components. Looking forward, the concept may inspire extensions to optical frequencies, dynamic reconfigurability, and broader exploration of non-Hermitian phenomena enabled by gauge-regulated loss[43-45].

# Methods

**Simulation:** We simulated the transmission and reflection spectra of the device using the commercial software CST Studio Suite. The relative permeability tensor of the gyromagnetic materials has the form $\tilde{\mu} = \begin{bmatrix} \mu_r & i\kappa & 0 \\ -i\kappa & \mu_r & 0 \\ 0 & 0 & 1 \end{bmatrix}$ , where $\mu_r = 1 + \frac{(\omega_0 + i\alpha\omega)\omega_m}{(\omega_0 + i\alpha\omega)^2 - \omega^2}$, $\kappa =$ $\frac{\omega\omega_m}{(\omega_0+i\alpha\omega)^2-\omega^2}$, $\omega_m = 4\pi\gamma M_s, \omega_0 = \gamma\mu_0 H_0$, and $\mu_0 H_0$ is the external magnetic field along the $z$ direction which is considered as 0.12T, $\gamma$ = 2.8MHz/Oe is the gyromagnetic ratio, α= 0.008 is the damping coefficient, and $\omega$ is the operating frequency. The saturation magnetization has been set to $4\pi M_s$=1780Oe. Permanent magnets are modeled as perfect electric conductors in simulations. Non-zero insertion loss is used in the simulation (PCB material) to match the practical experiment. Since the metasurface is periodic, we simulate a single unit cell with floquet boundary condition.

**Experimental setups:** During the near-field measurements, the sample was positioned horizontally between the excitation antenna and the scanning probe. A horn antenna placed below the sample was used to illuminate the metasurface with an approximately plane wave. The local $E_z$ field above the sample was detected by a dipole probe oriented along the same direction. The probe was fabricated from a coaxial cable by removing a 5 mm section of the outer conductor and exposing the central conductor, allowing the amplitude and phase of the electric field component parallel to the probe to be recorded. The probe was mounted on a three-axis motorized translation stage to map the spatial field distribution over the sample surface. For the far-field scattering measurements, the transmission and reflection spectra were obtained using two identical linearly polarized rectangular horn antennas connected to a vector network analyzer, with the frequency swept from 8 to 15 GHz.

## Acknowledgments

This work was supported by the New Cornerstone Science Foundation, the Research Grants Council of Hong Kong (AoE/P-502/20 and 17309021).

## Author contributions

S.Z. and Q.D.Y. conceived the idea. Q.D.Y designed the structure and performed numerical simulations. Q.D.Y., X.H.W., Z.F.L. and O.B.Y. performed theoretical analysis. Q.D.Y. and X.H.W. conducted the experiments under the supervision of Y.W. and S.Z. Q.D.Y. and Z.F.L. processed the experimental data. Q.D.Y., X.H.W., Z.F.L. and analyzed the experimental results. Q.D.Y. wrote the manuscript with input from all authors. S.Z. supervised the project. All authors contributed to the discussion.

## Competing interests

The authors declare no competing interests. Data and materials availability: All data are available in the manuscript or the supplementary materials. All data, code, and materials in the main text or the supplementary materials are available from Shuang Zhang.

# Figures

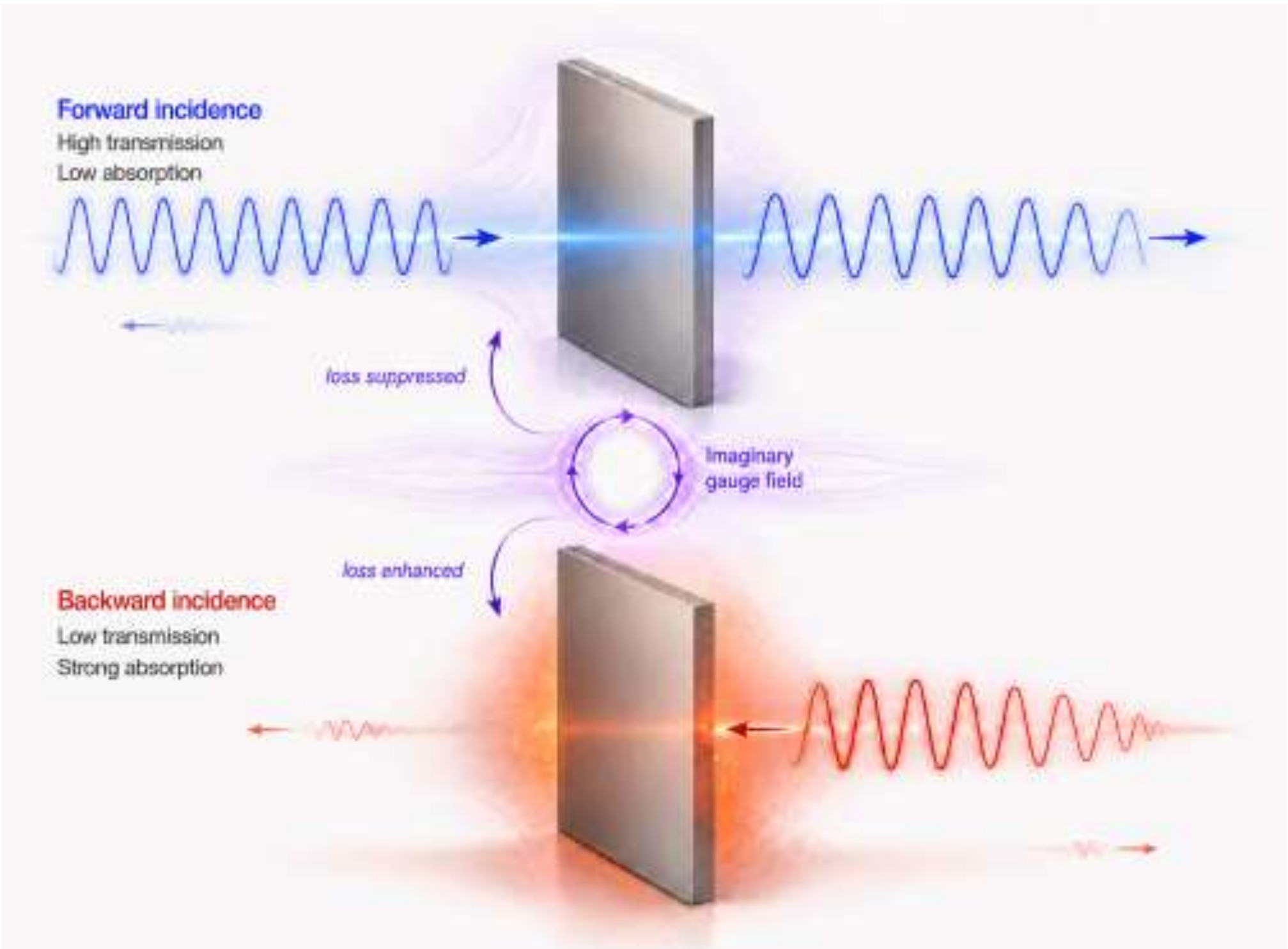

**Fig. 1 | Imaginary gauge strategy for direction-dependent loss.** Schematic of the imaginary gauge design principle. Introducing an artificial gauge potential makes the propagation factor path dependent. The imaginary gauge potential ($A_z''$) reduces the net attenuation accumulated by the forward channel for forward incidence. Reversing the propagation direction reverses the gauge accumulation, so the same $A_z''$ enhances the attenuation of the backward wave. Thus, the forward wave is transmitted with weak dissipation, whereas the backward incident wave is absorbed inside the metasurface.

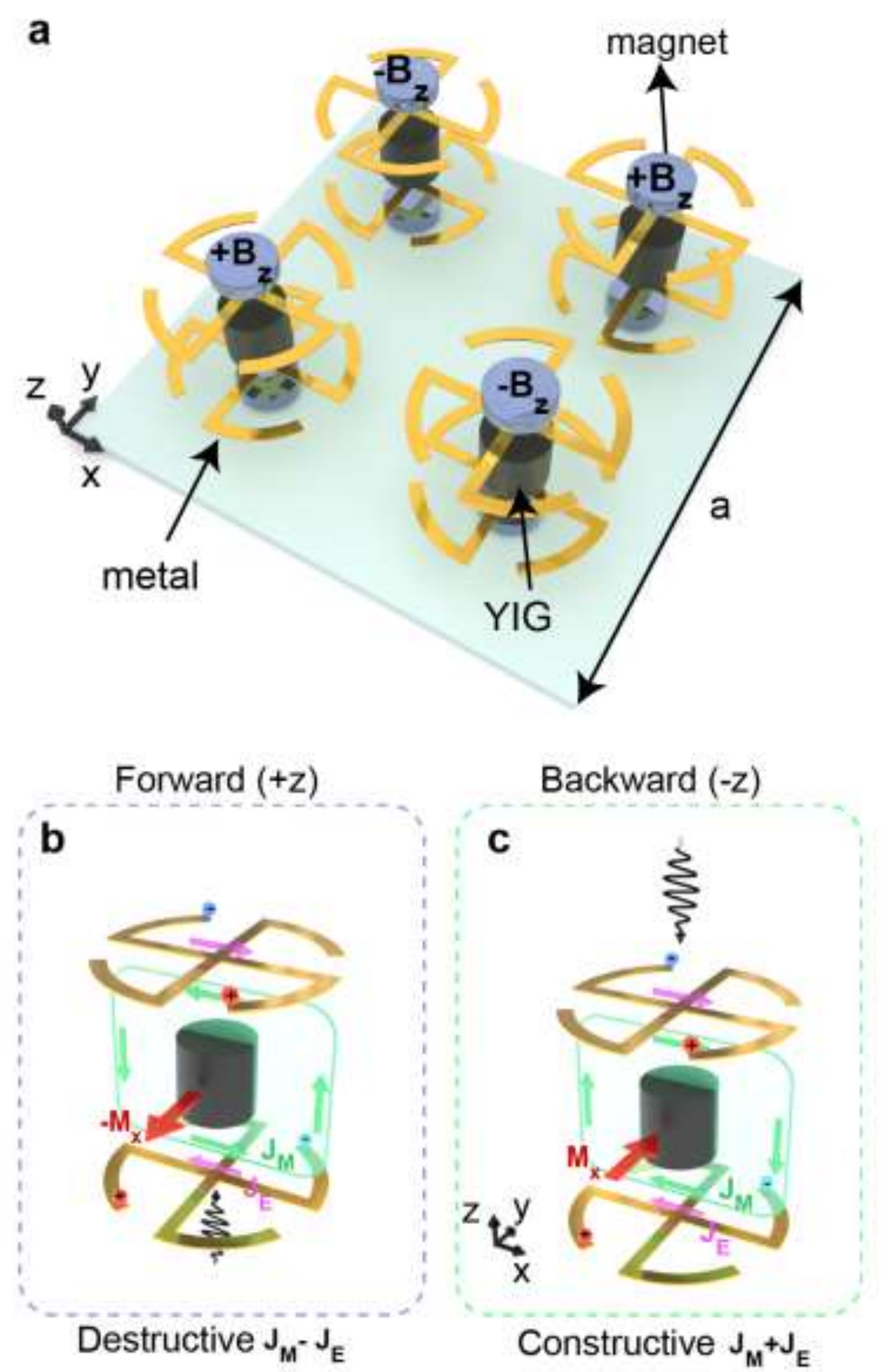

**Fig. 2 | Metasurface implementation and microscopic origin of the dissipative gauge field. a**, Geometry of the moving-type bianisotropic metasurface. The unit cell consists of four metamolecules composed of YIG rods, permanent magnets and planar metallic resonators, a=24 mm. The direction of local static magnetic field ($\pm B_z$) is determined by direction of magnets. **b**, Microscopic current picture for forward incidence. The incident electric field ($E_x$) drives charge accumulation at the metallic arms and excites a direct resonant current ($J_E$). Simultaneously, the magnetic field ($H_y$) generates a magnetic dipole ($M_x$) by gyromagnetic response of YIG and induces a secondary current ($J_M$) through Faraday induction. the two currents contribution interfere destructively in the relevant resonant arms. **c**, For backward incidence, the electric-field-driven current keeps the same reference phase, whereas the magnetic-field-driven current reverses sign because $H_y$ and $M_x$ change sign when the propagation direction is reversed. The two currents $J_E$ and $J_E$ add constructively.

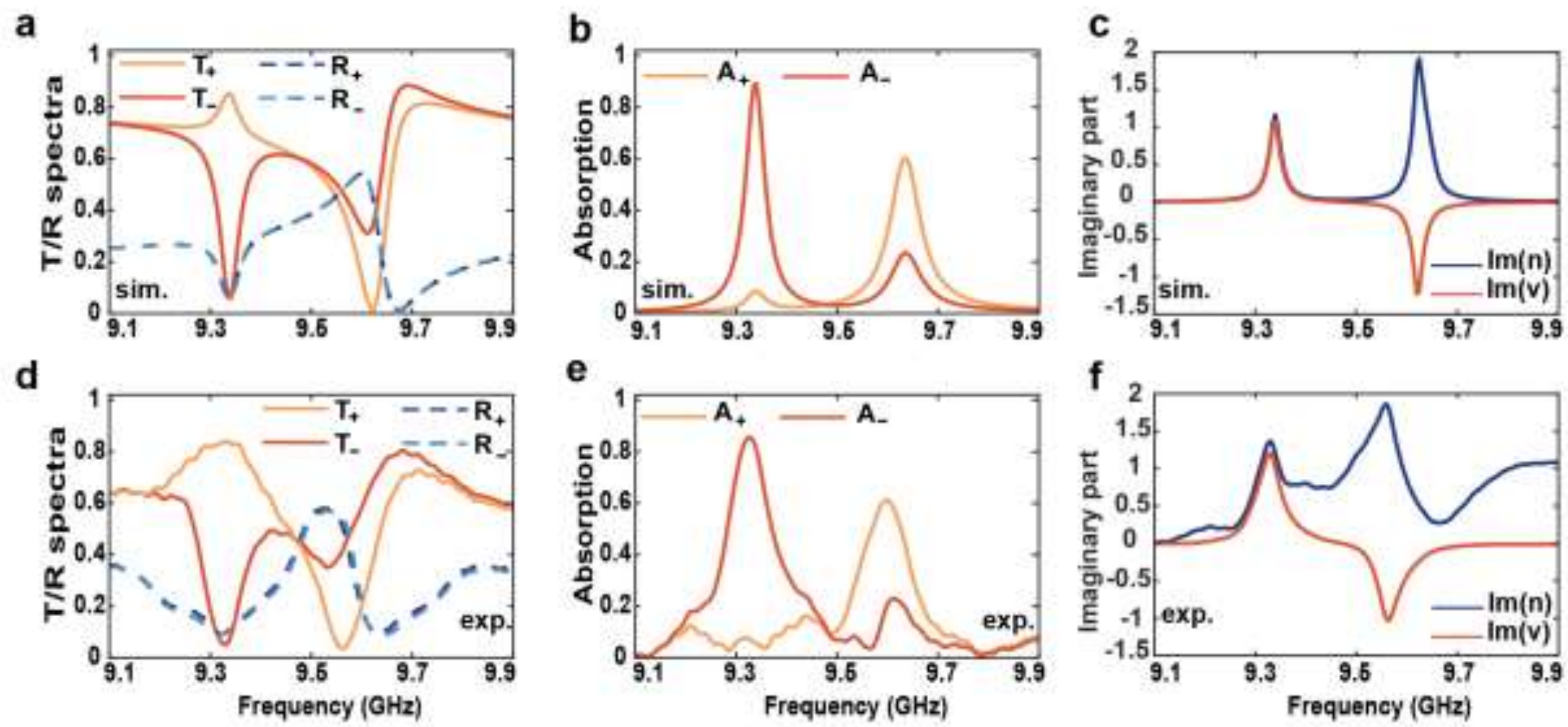

**Fig. 3 | Simulated and measured nonreciprocal transmission and absorption.** Simulated **(a)** and measured **(d)** co-polarized transmission and reflection spectra for waves incident from the forward and backward directions. Simulated **(b)** and measured **(e)** absorbance spectra calculated from the scattering coefficients. The labels "T", "R" and "A" indicate transmission, reflection, and absorption respectively. The labels "+" and "-" represent forward and backward incident directions respectively. Simulated **(c)** and measured **(f)** retrieved effective loss parameters. The imaginary part of $n$ and $v$ represent the ordinary loss background and the dissipative gauge component, respectively.

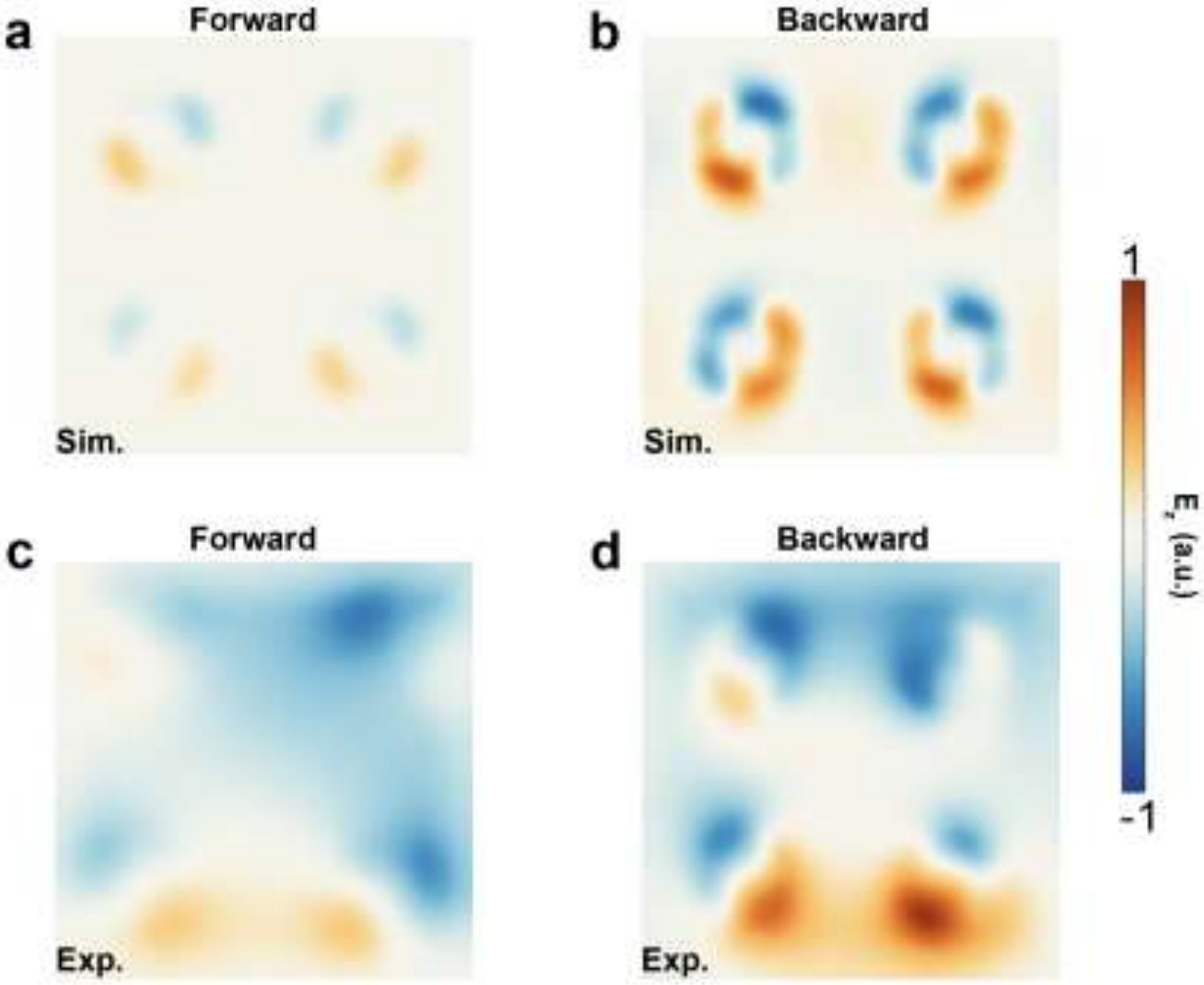

**Fig. 4 | Near-field evidence of gauge-controlled dissipative excitation. a**, Simulated surface $(E_z)$ distribution for forward incidence at the 9.32 GHz in one unitcell. The weak $(E_z)$ hot spots indicate suppressed surface charge accumulation and weak excitation of the dominant resonant current mode. **b**, Simulated surface $(E_z)$ distribution for backward incidence. The enhanced $(E_z)$ hot spots near the resonator gaps and metallic arms indicate stronger charge accumulation, enhanced resonant current excitation and larger dissipated power. Measured surface $(E_z)$ map for forward incidence **(c)** and backward incidence **(d)**.